\documentclass[prm,twocolumn,preprintnumbers,amsmath,amssymb,floatfix]{revtex4-1}
\usepackage{graphicx}
\usepackage{epstopdf}
\usepackage{float}
\usepackage{placeins}
\usepackage{hyperref,graphicx}
\usepackage{dcolumn}
\usepackage{bm}
\usepackage{color}
\usepackage{natbib}
\usepackage{amsmath}
\usepackage{amssymb}
\usepackage{multirow} 
\usepackage{hyperref}
\hypersetup{colorlinks=true, linkcolor=blue, citecolor=red, urlcolor=magenta, pdftitle={2DMOF}, pdfauthor={Saurabh Ghosh}}
\begin{document}
\begin{center}
\begin{large}
\end{large}
\end{center}
\title{Engineering strong magnetoelectricity using a hexagonal 2D material on electron-doped hexagonal LuFeO$_3$}
\author{M. J. Swamynadhan$^1$}
\author{Andrew O'Hara$^2$}
\author{Saurabh Ghosh$^{1}$}
\email {saurabhghosh2802@gmail.com}
\author{Sokrates T. Pantelides$^{2}$}
\email {pantelides@vanderbilt.edu}
\affiliation{$^1$Department of Physics and Nanotechnology, SRM Institute of Science and Technology, Kattankulathur - 603 203, Tamil Nadu, India}
\affiliation{$^2$Department of Physics and Astronomy, Vanderbilt University, Nashville, Tennessee, 37235, USA}

\begin{abstract}
Cubic perovskite-structure ABO$_3$ and A$_{1-x}$A$^{\prime}$$_x$BO$_3$-type oxides have been investigated extensively while their hexagonal-structure versions have received minimal attention, even though they are multiferroic and can form heterostructures with the manifold hexagonal two-dimensional materials. 
Hexagonal ferrites of the form RFeO$_3$, where R is yttrium or a rare-earth element such as Lu, Yb, etc., feature coupled ferroelectricity (FE) and weak-ferromagnetism (wFM), exhibiting linear magnetoelectricity. 
Their only drawback is the weak-ferromagnetism. 
In this paper, we employ density-functional-theory (DFT) calculations on hexagonal LuFeO$_3$ ($h$-LFO), targeting its magnetic ordering by electron doping anticipating spin-disproportionation of the Fe sublattices. 
Indeed, we show that spin-disproportionation in heavily-electron-doped versions Lu$_{1-x}$Hf$_x$FeO$_3$ ($h$-LHFO), especially for x=1/3 and 1/2, leads to robust out-of-plane collinear ferrimagnetism that is stable at room temperature. 
Furthermore, the robust ferroelectricity of $h$-LFO persists via a Jahn-Teller metal-to-insulator transition. 
Finally, we construct a $h$-LHFO/$h$-2D heterostructure, where $h$-2D stands for the FE/FM monolayer MnSTe, and demonstrate strong magnetoelectric coupling, namely manipulation of magnetic skyrmions in MnSTe by an external electric field through the $h$-LHFO polarization, opening up a new realm for magnetoelectric applications. 
\end{abstract}
\maketitle
\section{Introduction}

For several decades, transition-metal oxides (TMOs) in the perovskite structure have been extensively studied in the prototype ABO$_3$ stoichiometry, and also in “doped" forms in which the A- or B-cation sublattice is shared by two different elements \cite{tao2009direct, 41luo2007orbital}. 
These materials have been studied as thin films and in heterostructures and superlattices, all of which often exhibit unique emergent electronic \cite{1berger2011band, 2wagner2019property, 3park2020first, malyi2023insulating}, magnetic \cite{4saghayezhian2019atomic, 5ghosh2015linear, 6ederer2008electric, ahmed2023effective}, electrical \cite{4saghayezhian2019atomic, 5ghosh2015linear, 6ederer2008electric}, optical \cite{3park2020first, 7zhou2015optical}, and vibrational \cite{8hoglund2022emergent, 9luo2015electronic} properties. 
Several such TMOs are used in diverse applications such as magnetic storage, ferroelectric memories, photovoltaics, etc.
\par
Several ABO$_3$-type TMOs, including rare-earth ferrites and manganites, also exist in hexagonal forms ($h$-ABO$_3$). 
These polymorphs, however, have not been investigated extensively, though they are intrinsically multiferroic, featuring magnetoelectricity, namely coupled magnetic ordering and ferroelectricity (FE)  \cite{10choi2010insulating, 11chae2010self, 12das2014bulk, 13disseler2015magnetic, 14wang2013room, 15moyer2014intrinsic}. 
This property enables switching the magnetization by an electric field, a key requirement for spintronic device applications \cite{16ramesh2010new, 17pantel2012reversible}. 
However, hexagonal TMOs typically feature weak ferromagnetism (wFM) from canted spins.
Only a few papers have explored possible emergent properties in heterostructures \cite{15moyer2014intrinsic, 22ikeda2005ferroelectricity, 23niermann2012dielectric, 24lafuerza2013intrinsic}
and superlattices \cite{21fan2020site, 25mundy2016atomically}.
Alloying of either of the cation sublattices has only been pursued with isoelectronic substitutions  \cite{18baghizadeh2021interplay, 19du2018vortex, 20naveen2019multiferroic}, i.e., the introduction of excess electrons or holes (“doping") has not been explored fully even though it has long been known to be an effective way to tune magnetic and other properties of perovskite-structure oxides  \cite{attfield2001structure, kim2018half} and interest in these materials has recently grown \cite{26nordlander2022epitaxy, 27kumar2022synthesis,das2023coupling}.
\par
The second major potential driver that remains to be explored is the fact that the vast majority of the hundreds of two-dimensional (2D) materials that have been discovered and studied since the advent of graphene have a hexagonal structure, i.e, 2D-on-oxides functional or even multifunctional heterostructures are possible. 
More specifically, the recent discovery of  FM \cite{28huang2017layer, 29gong2017discovery, 30deng2018gate}, FE \cite{31liu2016room, 32xiao2018intrinsic, 33yuan2019room}, and multiferroic  2D materials triggered theoretical interest in fully 2D FM/FE heterostructures \cite{34seixas2016multiferroic, 35jin2022designing}for the generation and manipulation of skyrmions and other interesting magnetic textures, possibly with the help of external magnetic fields. 
Clearly, multiferroic hexagonal TMOs are a promising arena to explore 2D-on-substrates magnetoelectric and other functional heterostructures.
\par
 In this paper, we employ density-functional-theory (DFT) calculations (see Methods) to demonstrate that i) electron doping of $h$-LuFeO$_3$ ($h$-LFO) 
 by Lu$\rightarrow$Hf substitution induces robust, room-temperature ferrimagnetic (FiM) ordering while retaining the intrinsic ferroelectricity and ii) by placing a 2D multiferroic layer on thusly doped $h$-LFO, skyrmions stabilized by a magnetic field can be "switched" by reversing an external electric field. 
 We first adopt the logic that doping with excess electrons can induce spin disproportionation in the Fe sublattices.
 Indeed, using DFT calculations, we find that n-electron doping by replacing n out of the six Lu atoms per primitive unit cell with Hf atoms, 
 resulting in $h$-Lu$_{1-x}$Hf$_x$FeO$_3$, where x=n/6, the density of states (DOS) projected on the 3$d$ orbitals of n of the six Fe atoms undergoes substantial redistribution and justifies a nominal Fe$^{+2}$ label, even though the electron redistribution in space is minimal and the change in the Fe magnetic moment is relatively small. 
 We also find that the electronic redistribution on the energy axis is triggered by a polar mode which is a combination of Jahn-Teller distortion \cite{38loa2001pressure, 39han2000metal, 40toulemonde1999changes} and the trimmer distortion \cite{14wang2013room, 15moyer2014intrinsic}.
 This mode, which we shall refer to as polar mode or polar distortion, causes a metal-to-insulator (MIT) transition, making doped hexagonal TMOs even more interesting in the hunt of designing polar metals and in studies of structurally triggered MIT transitions. 
 In particular, along the polar pathways for different n values, we detect the transient presence of polar insulating, polar zero-band-gap semiconducting, and polar metallic phases that can in principle be stabilized by strain. 
 Furthermore, various spin configurations can be created by tuning x, namely the Fe$^{+2}$/Fe$^{+3}$ ratio. Among them, Lu$_{2/3}$Hf$_{1/3}$FeO$_3$ and Lu$_{1/2}$Hf$_{1/2}$FeO$_3$ show large, spontaneous, collinear out-of-plane magnetization that persists at room temperature, while Lu$_{1/3}$Hf$_{2/3}$FeO$_3$ features almost collinear FiM even at 0 K (in all cases, magnetic coupling between the layers is weak, but it can be enhanced by applying strain; overall FM should be easily achievable by a small external magnetic field). 
 Furthermore, $ab initio$ molecular dynamics (AIMD) simulations reveal that Fe planes that have both Fe$^{+2}$ and Fe$^{+3}$ sublattices feature collinear out-of-plane FiM ordering. 
 The feasibility of forming desirable compositions is discussed regarding minimum-energy chemical reaction pathways. 
 Finally, we place a monolayer of FM/FE MnSTe on $h$-Lu$_{1/3}$Hf$_{2/3}$FeO$_3$ and demonstrate that skyrmions can be generated and stabilized at different sizes by a relatively weak external magnetic field and can be switched by an external electric field through the coupling of the magnetizations and polarizations in the two materials. 

\section{Computational Details}
Density functional theory (DFT) \cite{50hohenberg1964inhomogeneous} calculations have been performed using the Vienna ab initio simulation package (VASP) \cite{vasphafner2008ab} within the choice of the projector augmented waves (PAW) basis set \cite{51kresse1996efficient, 52blochl1994projector}. 
The Perdew-Burke-Ernzerh (PBE) functional has been used to treat the exchange and correlation functional \cite{53perdew1996generalized}. 
The cutoff energy is set as 520.0 eV, and a k-point mesh of 4×4×2 has been used for the calculations. 
All the structures have been fully relaxed until all forces on all atoms are smaller than 0.001 eV/$\AA$. We considered Lu-$4f$ states in the core, and for Fe-$3d$ states, we chose effective Hubbard correction U-J$_H$ = 4.5 eV.
We arrived at the ground state spin configuration of $h$Lu$_{1-x}$Hf$_x$FeO$_3$ by incorporating spin-orbit coupling and relaxing the spin and geometry starting from the magnetic configuration of the parent system pure $h$-LFO (see supplementary information for more details). 
The Symmetric exchange interaction (J's), Antisymmetric exchange interaction (DMI's), and easy axis are calculated using the method proposed by Xiang et al. in ref. \cite{xiang2011predicting} and Weingart et. al. in ref., \cite{weingart2012noncollinear}. 
For AIMD simulations, the canonical (NVT) ensemble has been used \cite{54nose1984unified, 55kresse1993ab, 56kresse1994ab} with a plane wave-basis cutoff energy of 450 eV with 0.5 fs time interval (time steps) between MD steps. 
The system was simulated below the curie temperature 100K for 10ps to obtain equilibrium conditions. The AMPLIMODES (symmetry mode analysis) \cite{57kroumova2001pseudo} and PSEUDO \cite{58kirov2003neutron} (a program for pseudosymmetry search) have been used to understand the structural modes involved in the phase transition.
We simulated the magnetic texture for the obtained magnetic parameters by solving the Landau–Lifshitz–Gilbert (LLG) using a spin dynamic code Spirit \cite{49muller2019spirit}

\section{Results and discussion}
%
\begin{figure*}%
\centering
\includegraphics[width=0.9\textwidth]{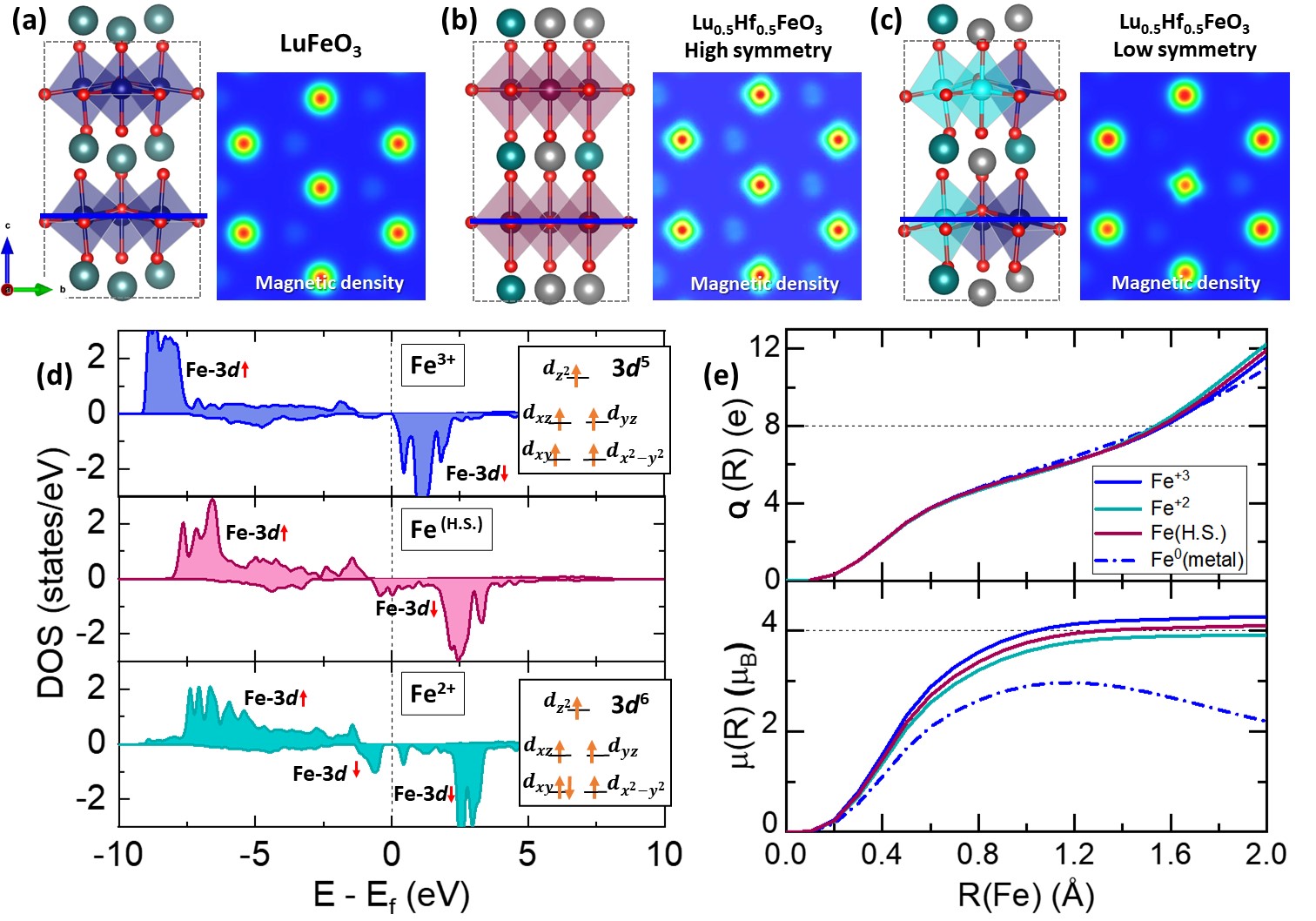}
\caption{Electron doping and oxidation-state ordering in $h$-LuFeO$_3$. (a) Side- and top-view schematics of the structure of $h$-LFO, (b) high-symmetry 50$\%$-doped $h$-LFO (Lu$_{1/2}$Hf$_{1/2}$FeO$_3$), and (c) relaxed distorted 50$\%$doped $h$-LFO. The dotted rectangles are the layer unit cells. Persian-green, grey, and red at-oms indicate Lu, Hf and O atoms. Purple and cyan polyhedra indicate Fe$^{+3}$ and Fe$^{+2}$, respectively. The top views are in the Fe plane and display the magnetic density in that plane. (d) DFT-calculated projected density of states (PDOS) of Fe$^{+3}$ in $h$-LFO (top panel), of Fe(HS) in undistorted 50$\%$-Hf-doped $h$-LFO (middle panel), and Fe$^{+2}$ in 50$\%$-Hf-doped $h$-LFO (bottom panel). Schematics in the insets show the Fe$^{+3}$ and Fe$^{+2}$ crystal-field splittings and occupancies as per the crystal-field/Hund’s-rule model). (e) Integrated spherically averaged electronic charge within radius R about Fe nuclei in different crystalline environments. The dotted lines denote the number of electrons in a neutral Fe atom (top) and the nominal magnetic moment for Fe$^{+2}$ (bottom). }\label{fig1}
\end{figure*}
$h$-LFO (space group P6$_3$cm), which has six nominal Fe$^{+3}$ ions per primitive unit cell, shown in Figure 1a(left), featuring trigonal bipyramids (TBP) BO$_5$ as basic building blocks \cite{10choi2010insulating, 11chae2010self, 12das2014bulk, 13disseler2015magnetic, 14wang2013room, 15moyer2014intrinsic}. 
The dotted rectangle denotes the primitive unit cell.
Ferroelectricity in $h$-LFO is of an improper nature \cite{36APLevanyuk_1974, 37benedek2022hybrid}, as the rare-earth Lu (Persian-green atoms) and oxygen (red atoms) buckle from their centrosymmetric plane to a polar structure. 
Below a critical ordering temperature Tc = 130 K, $h$-LFO also features wFM that arises from the slight canting of A$_2$ noncollinear Fe spins (see Supplementary Figure S1 for A$_2$ and other possible spin states in hexagonal TMOs) from in-plane to the out-of-plane direction \cite{12das2014bulk, 14wang2013room, 15moyer2014intrinsic}.
\par
Conventional analysis based on nominal oxidation states, Hund’s rule, and crystal-field theory provides initial insights as to what would happen when we electron dope $h$-LFO. 
In pure $h$-LFO the oxidation states are Lu$^{+3}$, Fe$^{+3}$, and O$^{-2}$; the Fe$^{+3}$ ions feature $d^{5\uparrow}$ magnetic configurations with crystal field-splittings as shown in the inset in Fig. 1d(top panel) and a magnetic moment $\mu$=5 $\mu_B$. 
If one out of the six Lu atoms in the unit cell is replaced by a Hf atom, presumed to enter as Hf$^{+4}$ ("one-electron doping" corresponding to x=1/6 in Lu$_{1-x}$Hf$_x$FeO$_3$), one assumes that the extra electron localizes on one of the six Fe$^{+3}$ atoms, converting it to Fe$^{+2}$ [$d^{5\uparrow1\downarrow}$] configuration and $\mu$=4 $\mu_B$ [see crystal-field splittings and occupations in the inset of Fig. 1d(bottom panel)]. 
One then might expect that the disproportionation into one Fe$^{+2}$ and five Fe$^{+3}$ sublattices may trigger spontaneous collinear FiM and hence FE/FiM multiferroicity as in the case of LuFe$_2$O$_4$ \cite{25mundy2016atomically}. 
Similar analysis applies to replacing 2, 3, 4, or 5 Lu atoms with Hf atoms per unit cell.  
\par
Density-functional-theory (DFT) calculations of pure and 50$\%$-doped $h$-LFO (see schematics of the high-symmetry undistorted and relaxed distorted structures in Figs. 1b and 1c, respectively) offer support for the above notions, but also find that the quantum physical reality is considerably more complex and at odds with the commonly used interpretation of oxidation states as physical charge of the ions. 
First, we note that the DOS projected on Fe$^{+3}$-3$d$ orbitals of $h$-LFO, shown in Fig. 1d(top), in-deed reveals almost total $d$-orbital spin polarization: the valence bands contain primarily spin-up electrons while the empty spin-down bands exist immediately above the Fermi energy, available to be occupied by doping electrons. 
In the case of three-electron-doped Lu$_{1/2}$Hf$_{1/2}$FeO$_3$, the DOS projected on the $d$ orbitals of three of the six Fe atoms remains relatively unchanged (not shown), preserving their Fe$^{+3}$ designation, while the DOS projected on the $d$ orbitals of the other three Fe atoms contain roughly one spin down electron below the Fermi energy [Fig. 1d(bottom)], signaling a conversion to Fe$^{+2}$ (see also Fig. S2 where we deduce the same inference from the energy bands).  
\par
\begin{figure*}%
\centering
\includegraphics[width=0.9\textwidth]{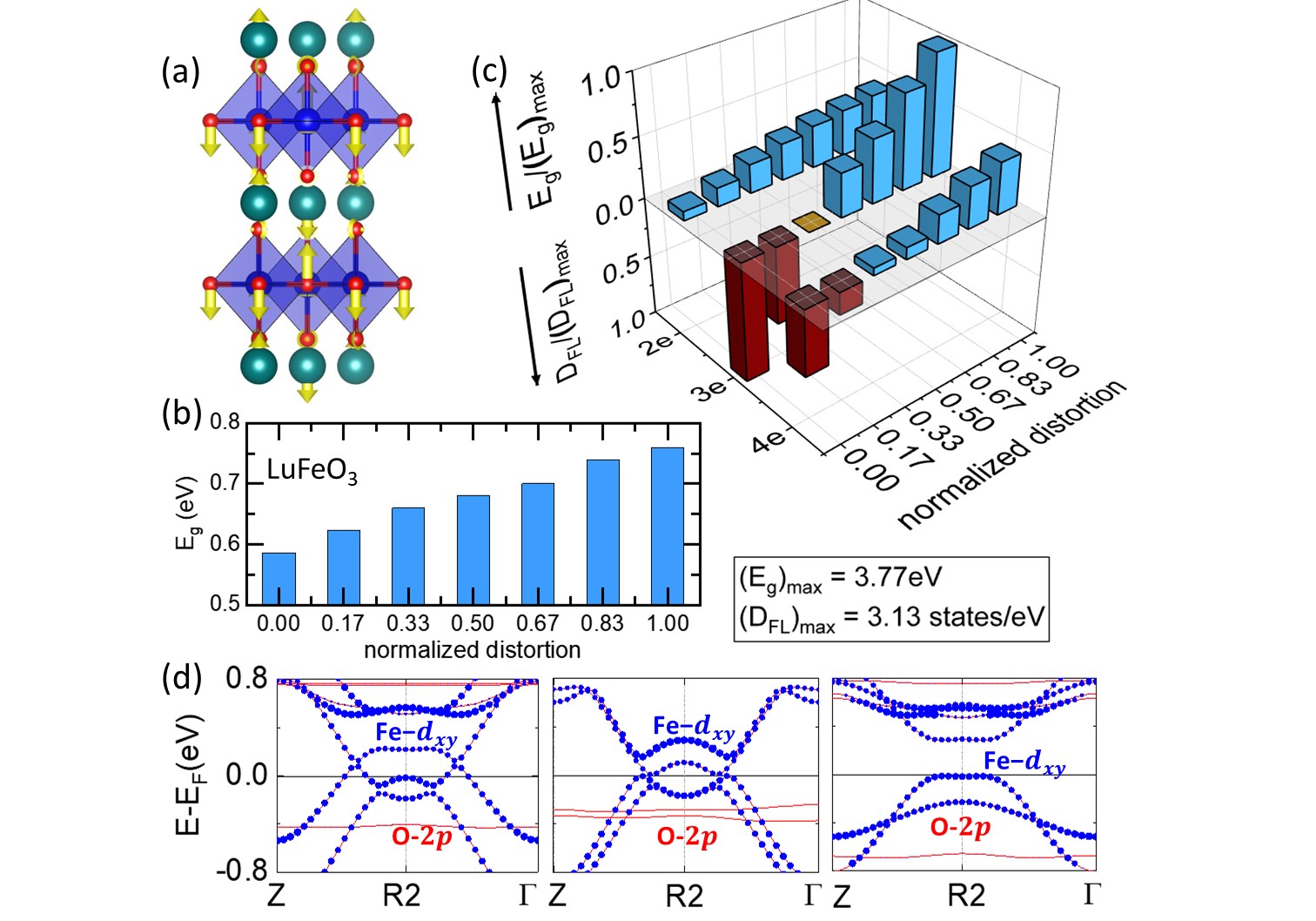}
\caption{Metal-to-Insulator transitions. (a) Coupled trimer distortion, BO$_5$ tilting mode. (b) Energy gap E$_g$ as a function of the polar-distortion amplitude for $h$-LFO. (c) E$_g$’s and densities of states at the Fermi level (DFL) as functions of the polar distortion in 2e$^-$, 3e$^-$, and 4e$^-$ doped systems. The values are normalized to the maximum E$_g$=3.77 eV and maximum DFL=3.13 states/eV of all three materials $d$ Band structures showing the (left) metallic phase in the NP state, (middle) band touching phase in the intermediate state and (right) insulating phase in the polar state of the 3e$^-$ doped system. These band structures help us understand the underlying mechanism.}\label{fig1}
\end{figure*}
\par
\begin{figure*}%
\centering
\includegraphics[width=1\textwidth]{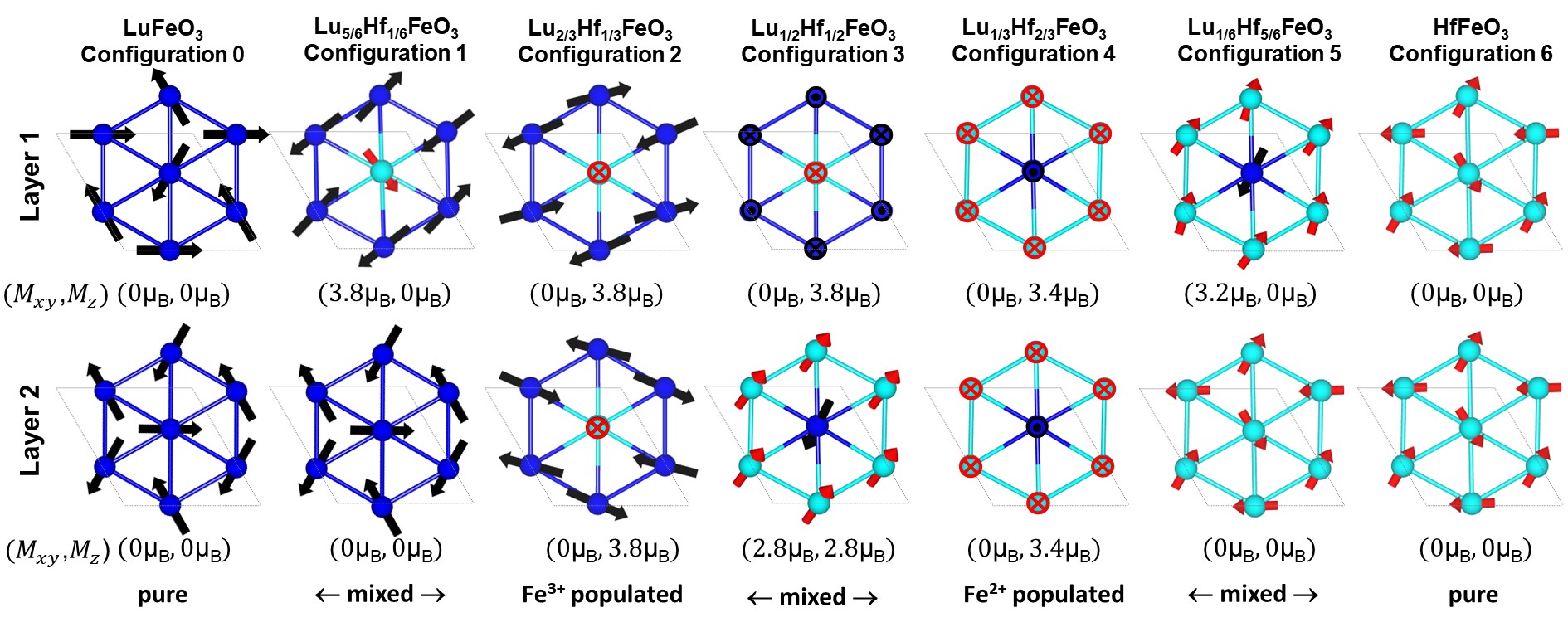}
\caption{Predicted magnetic configurations of pure and electron-doped $h$-LFO at T=0 K. Blue and cyan atoms represent Fe$^{+3}$ and Fe$^{+2}$ respectively. Black and red arrows indicate the magnetic spin orien-tation of Fe$^{+3}$ (~4.2 $\mu_B$) and Fe$^{+2}$ (~3.7 $\mu_B$) respectively. Circled cross and circled dot represent into the plane and out of plane vectors, respectively.}\label{fig1}
\end{figure*}
We note, however, that though the existence of two distinct types of Fe sublattices is definite, the precepts of oxidation-state-based analysis have limited realization in the quantum regime. If, upon electron doping, we do not allow polar distortions to occur, there is only minimal disproportionation among the six Fe atoms [the projected DOS is shown in Fig. 1d(middle)]. 
Upon allowing the polar distortion, Fig. 1d(bottom) shows the opening of a gap at the Fermi energy (metal-to-insulator transition) that can be traced to a linear Jahn-Teller effect \cite{38loa2001pressure, 39han2000metal, 40toulemonde1999changes}. 
Furthermore, Figs. 1d(top) and 1d(bottom) show that doping induces a very substantial redistribution of the d orbitals on the energy axis (see Fig. S3 for a more detailed analysis). 
\par
In order to examine changes in the physical electron charge in going from  Fe+3 to Fe+2, we examine the spherically averaged integrated electron density around the Fe nucleus, defined by Q(R)$=4\pi\int_{0}^{R}\rho(r)r^2 dr$, where the integral is carried out within concentric spheres of increasing radius R \cite{41luo2007orbital}. 
Q(R) curves for Fe$^{+3}$,  Fe$^{+2}$, Fe(HS) and Fe$^0$ for metallic iron are shown in Fig1e(top).
It is clear that Q(R) is essentially identical for all four Fe forms. 
There exists no value of the radius R at which there is a discernible $\Delta$Q between the curves of Fe$^{+3}$,  Fe$^{+2}$, Fe (HS), and the curve for Fe$^0$. 
In other words, the oxidation-state nominal charge of ionic crystals does not correspond to the physical electron distribution, as first discussed in 2007 \cite{41luo2007orbital} and 2008 \cite{42raebiger2008charge}, and all atoms are effectively always neutral \cite{41luo2007orbital}. 
In other words, the disproportionation of the Fe sublattices into Fe$^{+3}$ and Fe$^{+2}$ does not occur by rearrangements of the spherically average electron charge in the physical space, but by rearrangements in the energy space as demonstrated by the changes in going from the Fe$^{+3}$ PDOS in Fig. 1d(top) to the Fe$^{+2}$ PDOS in Fig. 1d(bottom). 
\par
The above results bear directly on the Fe magnetic moments that are a central focus of this paper as follows. 
Typically, DFT-based calculations of the magnetic moment of atoms in crystals quote a number obtained by the integral $\mu$(R)$=4\pi\int_{0}^{R}[\rho^{\uparrow} (r)-\rho^{\downarrow}(r)]r^2 dr$ for a fixed value of R, usually defined by the computer code. 
Figure 1d(bottom) shows the effect of the substantial redistribution of the Fe-$d$ orbitals that is evident in the three panels of Fig. 1a on the corresponding spherically averaged integrated net spin density $\mu$(R) around each distinct Fe nucleus defined by the above formula.
In Fig. 1e(bottom), we show that the integrated magnetic moment $\mu$(R) saturates to a unique value at large R. 
For the Fe species of interest here, Fe$^{+3}$,  Fe$^{+2}$, and  Fe(HS), these values are 4.3 $\mu_B$, 3.9 $\mu_B$, and 4.1 $\mu_B$. The trend from Fe$^{+3}$ to Fe$^{+2}$, namely 4.3 to 3.9 $\mu_B$ is indeed consistent with the oxidation-state-based nominal values of 5 and 4 $\mu_B$, respectively, but the numerical agreement is limited, reflecting the fact that the overall electron distribution around Fe$^{+3}$ and Fe$^{+2}$ atoms is essentially identical [Fig. 1e(top)]. 
Similarly, the delocalized part of the excess three 3$d$ electrons per unit cell before the polar distortions raise the saturation value of the Fe (HS)’s magnetic moment by only a small fraction of an electron and is almost invisible in the Q(R) plot in which all the valence electrons are included. 
Figures 1a, b, and c offer a visualization of the localized and delocalized spin densities in the three types of Fe. Note that the delocalized charge (shown in Figure S3) is minimal in all cases, consistent with the curves in Fig. 1e. 
\par
We now turn to the electrical and magnetic properties of $h$-LFO and its doped versions. 
$h$-LFO is inherently FE. 
To understand the effect of the polar distortion on the electronic structure and oxidation-state ordering, we considered the high-symmetry (HS) structures of $h$-LFO, Lu$_{2/3}$Hf$_{1/3}$FeO$_3$, Lu$_{1/2}$Hf$_{1/2}$FeO$_3$ and Lu$_{1/3}$Hf$_{2/3}$FeO$_3$ systems and performed electronic-structure calculations by moving all atoms in concert, as shown by the yellow arrows in Fig. 2a, freezing the low-symmetry (LS) polar distortions at various magnitudes. 
In the case of $h$-LFO, a 0.59 eV energy gap (Eg) in the HS structure increases gradually to 0.76 eV as the amplitude of the polar distortion reaches its maximum [Fig. 2b]. 
Figure 2c shows the effect of the polar distortions on the electronic structure in the doped materials. We plotted Eg for insulating phases and the density of states (DOS) at the Fermi level (DFL) for the metallic phases within a single frame. 
The z-axis in Figure 2c represents the normalized Eg (blue bars) for insulating phases and the normalized DFL (brown bars) for metallic phases.
In Lu$_{2/3}$Hf$_{1/3}$FeO$_3$, namely two-electron doping, we observe the same behavior as in $h$-LFO, but with an Eg that increases from $\sim$0.3 eV to $\sim$1.5 eV. 
On the other hand, Lu$_{1/2}$Hf$_{1/2}$FeO$_3$ (3-e doping) starts out as a metal in the HS state, but the polar distortion induces a gradual decrease in the DFL and, after a point of zero Eg, the material becomes insulating. 
In other words, a metal-to-insulator transition is induced by a polar distortion. 
A similar trend is observed in Lu$_{1/3}$Hf$_{2/3}$FeO$_3$ (4-e doping).
\par
The metallic, zero-band-gap, and insulating (just after opening the gap and not the ground state) band structures of Lu$_{1/2}$Hf$_{1/2}$FeO$_3$ (3-e doping) are shown in Figure 2d. 
In the terminal insulating band structure, the valence-band maximum is found to be composed of Fe$-3d$ and O$-2p$ states, as in the case of the parent compound \cite{43holinsworth2015direct}. 
In the 50$\%$-doped case, the extra electrons introduced by Hf substitution of Lu atoms make the system metallic in the HS state, but then a first-order Jahn-Teller (JT) distortion drives the polar distortion as discussed in Ref. \cite{38loa2001pressure, 39han2000metal, 40toulemonde1999changes}. 
This JT effect is verified by measuring the bond lengths between the atoms. 
The in-plane Fe$^{+2}$-O bond length elongates by 7$\%$ while the Fe$^{+3}$-O bond length shrinks to 3$\%$. 
As the Fe$^{+2}$-O bond length is elongated, the energy of the doublet state e$^{\prime\prime}$ of Fe$^{+2}$ comes down due to reduced Coulombic repulsion, which is observed in the band structure shown in Fig. 2d. 
In the ground state, $h$-LFO and its Hf-doped versions are insulating due to polar distortion and spin-orbit coupling. 
In the HS structures, all Fe atoms are identical, i.e., there are no Fe$^{+2}$ and Fe$^{+3}$ sublattices, while the minimally delocalized Fe-$3d$ electrons are somewhat enhanced (Fig. 1b). 
When the polar distortion is “switched on", BO$_5$ tilting, and the enlargement of BO$_5$ polyhedra lift the degeneracy and bring the Fe$-3d_{xy}$ down as shown in Figure2d(bottom).
Thus, the polar distortion mode drives the MTI transition.
The MTI transition happens at an appropriate "coupling strength" between the trimer distortion and the JT mode. 
This phenomenon occurs irrespective of the doping positions (Fig. S4).
\par
We turn now to magnetic properties. 
Aiming to achieve collinear FM or FiM in FE h-Lu$_{1-x}$Hf$_x$FeO$_3$, we first determined the magnetic ground state at Hf concentrations x= n/6, n=0 to 6, corresponding to n excess electrons per unit cell. 
In each case, we started with the noncollinear A$_2$ magnetic configuration for several Hf sites (see Fig. S4) in the unit cell and relaxed both the spins and the geometry in the presence of spin-orbit coupling (see Fig. S5). 
The relaxed magnetic con-figurations for the lowest-energy structures are indicated in the schematics of Fig. 3.
\par
We determined the major symmetric J$_{ij}$ and antisymmetric Dzyaloshinskii-Moriya (DM) exchange interactions D$_{ij}$ between nearest-neighbor Fe atoms and the magnetic easy axis at T=0 K to gain insights into the obtained relaxed spin configurations.
The calculated Js and DMIs are tabulated in Table S1. 
In pure LFO and HFO, all Fe sublattices are equivalent. 
The respective calculated Fe$^{+3}$-Fe$^{+3}$ and Fe$^{+2}$-Fe$^{+2}$ Js (see supplementary information for more details) reveal   AFM super-exchange interaction, which would create magnetic frustration in these triangular-lattice systems. 
Hence, these materials retain the A$_2$ state, as shown in Fig 3, configurations 0 and 6, respectively.
In the Fe$^{+3}$:Fe$^{+2}$ = 2:1 case of Lu$_{2/3}$Hf$_{1/3}$FeO$_3$, configuration 2 in Fig. 3, each Fe atom in the Fe$^{+2}$ sublattice is surrounded by six Fe$^{+3}$ irons. 
Here, the strong symmetric AFM exchange interaction between Fe$^{+3}$-Fe$^{+3}$ pairs at the vortices of the hexagonal ring leads to AFM oriented Fe$^{+3}$ spins. 
Due to strong DMI and the crystallographic c-axis being the easy axis, the Fe$^{+2}$ spins are oriented along the c-axis perpendicular to Fe$^{+3}$ spins to suppress frustration. 
Hence, by electron doping, we obtain large out-of-plane magnetization as large as $\sim$3.8 $\mu_B$ per three Fe atoms ($\sim$1.3 per Fe atom) in a FE $h$-LFO system. 
In the reciprocal case with Fe$^{+3}$:Fe$^{+2}$ = 1:2, namely configuration 4 with four excess electrons per unit cell, each Fe$^{+3}$ atom in each of the two Fe$^{+3}$ sublattices is surrounded by six Fe$^{+2}$ atoms. 
Once more, the relative strengths of the Js and DMIs account for the resulting collinear magnetism to avoid frustration. 
The magnetization of the other configurations can be understood similarly.
\par
In all the configurations of Fig. 3, inter-layer interactions are very weak, and, as a result, inter-layer FM and AFM configurations are essentially degenerate. 
By applying uni-axial strain along the crystallographic c-axis, the inter-layer distance can be reduced, thereby, we can tune the interaction between the successive layers (see more information in SM Fig S6). 
\par
So far, we have discussed the magnetic configurations of the magnetic ground states at 0 K. 
In order to determine the magnetic configurations at higher temperatures, we performed ab-initio molecular-dynamics (AIMD) simulations with SOC turned on. 
These simulations allow us to track the non-collinear magnetic ordering at each time step. 
We first performed simulations for the parent compound, $h$-LFO, to validate the technique at 100 K, below T$_N$ = 130 K. 
We confirmed that the system is stable in its A$_2$ configuration (shown in Figure S7). 
We then performed AIMD simulations at 100 K for all the doped configurations of Fig. 3, tracking the system's net M$_x$, M$_y$, and M$_z$ magnetization components for each time step. 
The results for Lu$_{5/6}$Hf$_{1/6}$FeO$_3$ are shown in Figure 4a. 
The system was started in the T = 0 configuration 1 of Fig. 3 with all spins in the xy plane. 
It is clear that, at 100 K, the system overcomes the DMI, and all the spins in the doped layer are aligned along the easy axis (crystallographic c-axis), while in the undoped layer, where all the Fe atoms are Fe$^{+3}$, the system maintains its in-plane magnetic state. 
Thus, the only contribution to the net magnetic moment per unit cell is from the Fe$^{+2}$, giving an out-of-plane magnetic moment of 3.8 $\mu_B$/unit cell. 
The net result is a perfect collinear FiM doped layer and a perfect non-collinear 120$^{\circ}$ oriented undoped layer, as shown in the inset of Figure 4a.
\par
\begin{figure}[h]%
\centering
\includegraphics[width=0.5\textwidth]{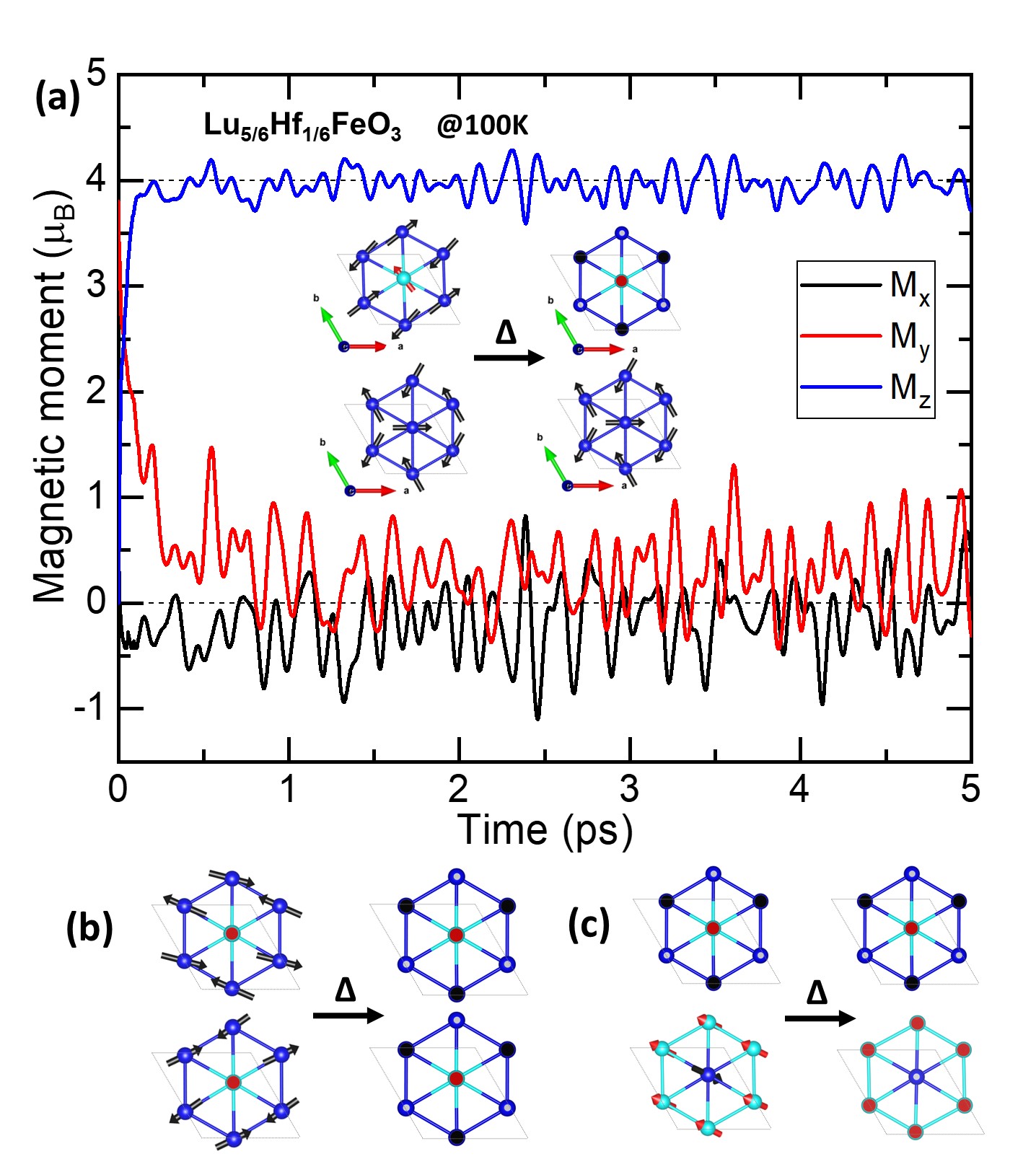}
\caption{Results of molecular-dynamics simulations. a) Magnetic moment as a function of simulation time at 100 K for the Lu$_{5/6}$Hf$_{1/6}$FeO$_3$ system (inset: change of magnetic configura-tion from in-plane to out-of-plane configuration). b) and c) the 100-K temperature effect on the magnetic configuration of Lu$_{2/3}$Hf$_{1/3}$FeO$_3$ and Lu$_{1/2}$Hf$_{1/2}$FeO$_3$, respectively.}\label{fig1}
\end{figure}
\begin{figure*}%
\centering
\includegraphics[width=0.9\textwidth]{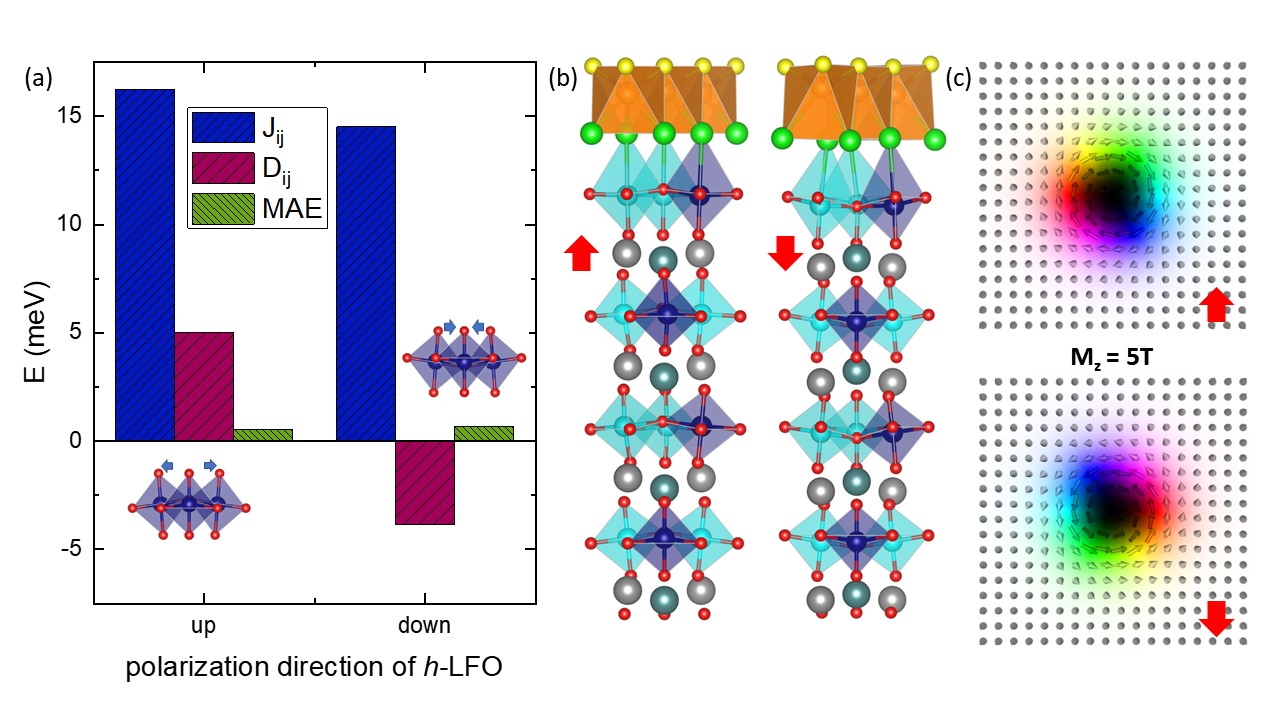}
\caption{Skyrmion generation and switching. a) The change in magnetic properties of monolayer MnSTe with respect to the polarization direction of $h$-Lu$_{1/3}$Hf$_{2/3}$FeO$_3$ (the inset figure shows the corresponding BO$_5$ tilting). b) Schematic showing the FE/FM monolayer MnSTe grown on $h$-Lu$_{1/3}$Hf$_{2/3}$FeO$_3$ whose polarization is along the crystallographic +c (left) and -c (right) axis respectively, c) The generated skyrmion on the FE/FM MnSTe layer in the presence of an external magnetic field H$_z$ =5T for polarization along +c (top) and -c (bottom) of $h$-Lu$_{1/3}$Hf$_{2/3}$FeO$_3$.}\label{fig1}
\end{figure*}
In the case of Lu$_{2/3}$Hf$_{1/3}$FeO$_3$, at 0 K, both the layers feature in-plane magnetic moments of 3.8 $\mu_B$. 
At 100K and 300K, both layers feature out-of-plane magnetic moment $\sim$1.33 $\mu_B$ per Fe atom as shown in Figure 4b, which is the largest magnetization we predict for Hf doped $h$-LFO   (the other MD simulations are shown in Figure S7 and S8).
The case of Lu$_{1/2}$Hf$_{1/2}$FeO$_3$ is shown in Fig. 4c At 0 K the Fe$^{+3}$-populated layer features out-of-plane magnetic moments of 3.8 $\mu_B$ while the Fe$^{+2}$ populated layer features in-plane 2.8 $\mu_B$. At 100 K both layers feature out-of-plane magnetic spins   $\sim$7 $\mu_B$/unit cell, 1.1 $\mu_B$/Fe. 
We observe similar behavior in all the concentrations. Whichever layer has two magnetic sublattices features collinear FiM along the crystallographic c-axis at 100 K. 
\par
Finally, we investigate the formation energies of h-Lu$_{1-x}$Hf$_x$FeO$_3$ with respect to the minimum energy reaction pathway by using the grand canonical approach combined with the linear programming problem (for more details, refer to the supplementary information) \cite{44r2007first, 45shaikh2021investigation}. 
The results show that the Hf-doped $h$-LFO systems are formable at ambient temperature and pressure. The formation energies are tabulated in SM Table 2.
\par
2D monolayer FM/FE MnSTe grown on top of h-Lu$_{1/3}$Hf$_{2/3}$FeO$_3$:
Finally, we placed an FM/FE monolayer MnSTe on h-Lu$_{1/3}$Hf$_{2/3}$FeO$_3$ to demonstrate a fruitful application on the generation of skyrmions. 
The 2D FM/FE choice is based on the previously reported detailed studies on intrinsic skyrmions in MnSTe \cite{46yuan2020intrinsic}.
As per previous reports, MnSTe has higher SOC energy and stronger DMI when compared to other Janus transition-metal dichalcogenides (TMDs) and tuneable intrinsic skyrmions with an external magnetic field \cite{46yuan2020intrinsic, 47liang2020very}. 
In h-Lu$_{1/3}$Hf$_{2/3}$FeO$_3$, both FeO and Lu/HfO$_2$ are stable terminations; their calculated surface energies \cite{48fredrickson2013wetting} are shown in Figure S9.
We have considered FeO termination of h-Lu$_{1/3}$Hf$_{2/3}$FeO$_3$ and Te layer for the heterojunction. 
When MnSTe is placed on top of the FeO layer of h-Lu$_{1/3}$Hf$_{2/3}$FeO$_3$, the Te of MnSTe completes the trigonal bipyramid, as shown in Figure 5b. 
As discussed earlier, the TBP tilt is coupled with the h-Lu$_{1/3}$Hf$_{2/3}$FeO$_3$ polarization. 
Thus, the Te completing the TBP also distorts according to the polarization direction, which results in a change in the direction of the DMI of MnSTe. 
The calculated magnetic parameters of MnSTe when placed on h-Lu$_{1/3}$Hf$_{2/3}$FeO$_3$ are plotted in Figure 5a. 
\par
To have a rough idea about the formation of skyrmions on the MnSTe surface, we simulated the magnetic texture for the obtained magnetic parameters by solving the Landau–Lifshitz–Gilbert (LLG) equations using the spin-dynamic code Spirit \cite{49muller2019spirit}. 
In the presence of an external magnetic field H$_z$=2 - 5T, skyrmions can be stabilized with a diameter ranging from $\sim$6 to $\sim$2 nm, respectively (shown in SI figure S10). 
The noteworthy point is that the skyrmion directions change when the direction of DMI changes, i.e., the skyrmion direction can be switched by switching the polarization direction of h-Lu$_{1/3}$Hf$_{2/3}$FeO$_3$ as shown in Figure 5c. 
Since our substrate, Hf doped $h$-LFO, has out-of-plane magnetization (M$_z$) to stabilize the skyrmion, and the magnetic strength of the substrate can be tuned by changing the concentration of Hf doping, this h-Lu$_{1/3}$Hf$_{2/3}$FeO$_3$/MnSTe heterostructure can be used to generate and tune skyrmions. 
The present findings provide a method to achieve strong magnetoelectric coupling in the 2D limit and a new perspective for designing skyrmion-based spintronics memory devices.

\section{Conclusion}
In conclusion, we have presented the fascinating results of electron doping of $h$-LuFeO$_3$. 
We described the general methodology to enhance the net magnetization in these materials by substitutional doping of tetravalent Hf in place of trivalent Lu.
We discussed the relative merits of the crystal-field/Hund's-rule model and the results of DFT calculations in understanding the effects of doping on the electron and the electron-spin distribution in these materials. 
We discussed the role of electrical polarization in inducing the differentiation of the equivalent Fe sublattices into two sublattices that their different effective oxidation states can label. 
The polarization distortion drives the doped material from insulating to metal or metal to an insulating state. 
Finally, we show that electron doping achieves an order of magnitude increase in magnetization compared to the parent $h$-LuFeO$_3$ material. 
The calculations indicate an almost collinear magnetic configuration in all the layers with non-equivalent Fe sublattices near room temperature. 
Further, we analyzed the strain effect on the inter-layer magnetic interactions. 
All the above-mentioned phenomena can be generalized to other hexagonal magnetic frustrated systems. 
The growth of hexagonal Hf-doped LuFeO$_3$ is expected to be practical, especially because our results do not depend on the dopant positions. 
Finally, we show a practical application for the large out-of-plane magnetization of these compounds to generate and tune skyrmions on a 2D monolayer FM/FE MnSTe, which may lead to novel applications in electronic and storage devices based on magnetoelectricity.

\section*{Acknowledgements}
M.J.S. and S.G. acknowledge DST-SERB Core Research Grands, File No. CRG/2018/001728 and  CRG/2018/004175 for funding the project. The authors thank the National Energy Research Scientific Computing Center, a U.S. Department of Energy Office of Science User Facility under Contract No. DE-AC02-05CH11231 for providing a supercomputer facility. Research at Vanderbilt University was supported by the U.S. Department of Energy, Office of Science, Basic Energy Sciences, Division of Materials Science and Engineering under Grant No. DE-FG02-09ER46554 and by the McMinn Endowment at Vanderbilt University






%


\end{document}